\documentclass[pss]{wiley2sp} 
\usepackage{amsmath}

\usepackage{amsmath,amssymb}
\usepackage{graphicx}

\usepackage[dvips,colorlinks=true,bookmarks=false,citecolor=blue,urlcolor=blue]{hyperref} 

\begin{document}
\title{Tunable hybrid surface waves supported by a graphene layer}
\titlerunning{Surface waves in graphene}

\author{
Ivan Iorsh\textsuperscript{\textsf{\bfseries \Ast ,1,2}},
Ilya Shadrivov\textsuperscript{\textsf{\bfseries 1 ,3}},
Pavel A. Belov\textsuperscript{\textsf{\bfseries 1 ,4 }},
and Yuri S. Kivshar\textsuperscript{\textsf{\bfseries 1 ,3}}}

\authorrunning{I. Iorsh et al.}

\mail{e-mail
    \textsf{i.iorsh@phoi.ifmo.ru}}

\institute{
    \textsuperscript{1}\,St.~Petersburg University of Information Technologies, Mechanics and Optics (ITMO), St. Petersburg 197101, Russia \\
      \textsuperscript{2}\,Department of Physics, Durham University, Durham, DH1 3LE, UK \\
    \textsuperscript{3}\,Nonlinear Physics Center, Research School of Physics and Engineering, Australian National University, Canberra ACT 0200, Australia \\
     \textsuperscript{4}\,Queen Mary, University of London, Mile End Road, London E1 4NS,UK}

\abstract{
\abstcol{
We study surface waves localized near a surface of a semi-infinite dielectric medium covered by a layer of graphene in the presence of a strong external magnetic field. We demonstrate that both TE-TM hybrid surface}{  plasmons can propagate along the graphene surface. We analyze the effect of the Hall conductivity on the dispersion of hybrid surface waves and suggest a possibility to tune the plasmon dispersion by the magnetic field.
}}


\maketitle

Graphene, two-dimensional lattice of carbon atoms, exhibits a wide range of unique electronic and optical properties~\cite{Rev1,Rev2,Rev3}. It was theoretically~\cite{theor2,theor1,theor3,theor4,theor5} and experimentally~\cite{Basovexp,Koppens_exp} demonstrated that specific type of localized waves, surface plasmon polaritons can propagate along a single layer of graphene or its bilayer. It was shown that both TM and TE polarized plasmons can exist in graphene~\cite{Newmode}, and their dispersion properties can be changed by applying an external gate voltage to the graphene sheet which allows to construct effective two-dimensional metamaterial structures based on graphene~\cite{Engheta}.

One of the key features of graphene is the linear dispersion of the electrons close to the band-edges which is similar to the dispersion of ultra-relativistic Dirac fermions~\cite{Rev1,Rev2,Rev3}. In particular, this leads to the square-root dependence of the electron cyclotron frequency on the magnetic field, and much larger separation of the Landau levels in graphene, and consequently to the possibility to observe quantum Hall effect at room temperatures~\cite{QHE}.

\begin{figure}[!h]
\centerline{\includegraphics[width = 0.65\columnwidth]{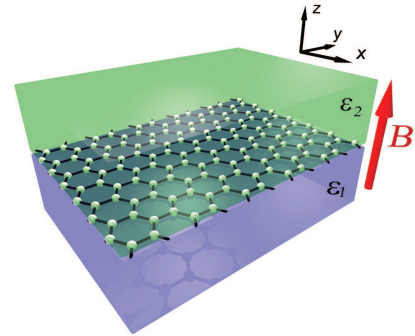}}
\caption{Geometry of the structure under consideration. A layer of graphene is placed at the interface of two dielectric media (in our case, the upper medium is air). External magnetic field is applied along the $z$ axis.}
\label{fig1}
\end{figure}

In this Letter, we study the properties of electromagnetic waves localized near a surface of a semi-infinite dielectric medium covered by a layer of graphene in the presence of a strong external magnetic field. We predict that the dispersion of the surface waves supported by a graphene layer can be tailored through the variation of the magnetic field.

We consider a simple geometry depicted schematically in Fig.~\ref{fig1}. A sheet of graphene is placed at an interface separating two dielectric media. The graphene layer is treated as a conductive surface~\cite{Newmode} defined by frequency-dependent conductivity $\sigma_0(\omega)$. When we apply a constant magnetic field perpendicular to the graphene sheet the conductivity becomes a tensor with components written as:
\begin{align}
\hat{\sigma}=\begin{pmatrix} \sigma_0 & \sigma_H \\ -\sigma_H & \sigma_0 \end{pmatrix},
\end{align}
where $\sigma_0$ and $\sigma_H$ are the longitudinal and Hall components of conductivity respectively.
   Boundary conditions for the tangential components of the electric and magnetic fields can be written in the form
\begin{align}
(\vec{E}_2-\vec{E}_1)\times \hat{z}=0,\quad(\vec{H}_2-\vec{H}_1)\times \hat{z}=\frac{4\pi}{c}\hat{\sigma}\vec{E}_{\|}.
\end{align}
For both media we are looking for the surface waves with harmonic temporal dependence $\mathrm{exp}(-i\omega t)$ and with spatial variation of the form:
$\vec{E}_{1,2},\vec{H}_{1,2}\sim \exp (i\beta x \pm \kappa_{1,2} z)$, where $\kappa_{1,2}=(\beta^2-\varepsilon_{1,2} k_0^2)^{1/2}$.

The waves in both the media can be presented as a linear combination of the TE and TM polarized waves, so that for the TM polarized waves we have:
\begin{align}
\vec{E_{1,2}}^{TM}=\left(\mp \frac{i\kappa_{1,2}}{k_0\varepsilon_{1,2}},0,-\frac{\beta}{k_0\varepsilon_{1,2}} \right);\vec{H_{1,2}}^{TM}=\left(0,1,0\right),
\end{align}
whereas for the TE polarized waves, we have 
\begin{align}
\vec{H_{1,2}}^{TE}=\left(\pm\frac{i\kappa_{1,2}}{k_0},0,\frac{\beta}{k_0} \right);\vec{E_{1,2}}^{TE}=\left(0,1,0\right).
\end{align}
Next, we consider linear combinations of TE and TM polarized waves in both the media, and write the electric fields in the form: $\vec{E_{1,2}}= \mathcal{A}\vec{E_{1,2}}^{TM}+\mathcal{B}\vec{E_{1,2}}^{TE}$. To match the waves at two different semi-finite spaces, we apply the continuity boundary conditions and derive
the resulting equations for the amplitudes $\mathcal{A}$ and $\mathcal{B}$,
\begin{align}
-\frac{4\pi\sigma_H\kappa_1}{ck_0\varepsilon_1}\mathcal{A}+\left(\frac{4\pi\sigma_0}{c}
-\frac{\kappa_2+\kappa_1}{k_0}\right)\mathcal{B}=0 \\
\left(\frac{4\pi i \sigma_0 \kappa_1}{c k_0 \varepsilon_1}+1+\frac{\kappa_1\varepsilon_2}{\kappa_2\varepsilon_1}\right)\mathcal{A}
-\frac{4\pi\sigma_H}{c}\mathcal{B}=0,
\end{align}
which allow to obtain the dispersion relations for the surface waves by setting the determinant of the system to zero,
\begin{align}
\left(\frac{i \sigma_0}{c}-\frac{\kappa_1+\kappa_2}{4\pi k_0}\right)\left(\frac{i \sigma_0}{c}+\frac{k_0\varepsilon_1}{4\pi\kappa_1}+\frac{k_0\varepsilon_2}{4\pi\kappa_2}\right)
= \frac{4\pi}{c^2}\sigma_H^2. \nonumber \\
\end{align}
When $\sigma_H$=0,  we have two solutions corresponding to the dispersion of both TE and TM surface waves
propagating in a graphene layer. However, when $\sigma_H \neq 0$ two polarizations are coupled and surface waves have hybrid TE-TM structure. Hall conductivity $\sigma_H$ plays a role of the coupling parameter.

\begin{figure}[!h]
\centerline{\includegraphics[width=0.9\columnwidth]{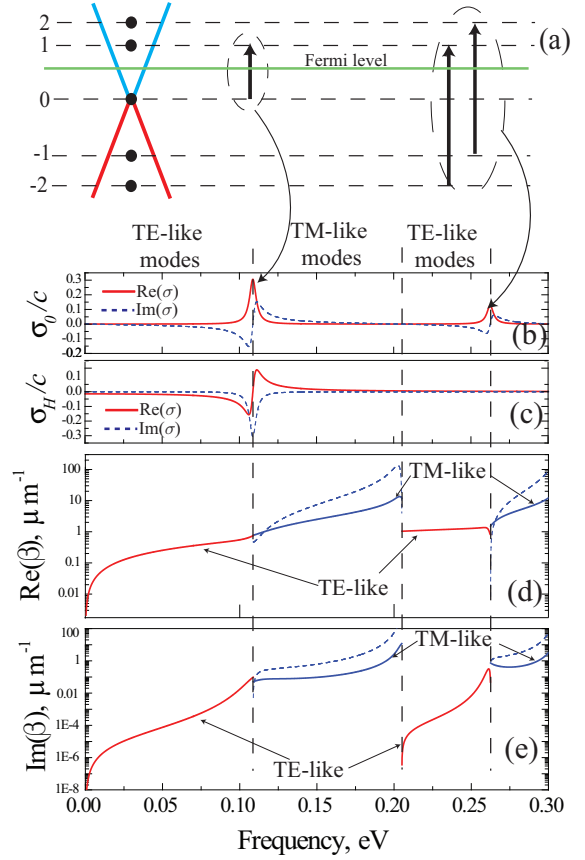}}
\caption{(a) First three low-energy transitions of the Landau levels in graphene. (b,c) Spectrum of the conductivities: red lines correspond to real part, blues lines - to imaginary part. (d) Dispersion of the real part of the waveguide number of the  surface states. (e) Dispersion of the imaginary part of the waveguide number. Solid red lines correspond to TE-like modes, solid blue to TM-like modes; dashed blue lines correspond to dispersion of TM-like modes without accounting for the Hall conductivity. }
\label{fig2}
\end{figure}

Expressions for the Hall and longitudinal conductivities can be obtained using the Kubo formula~\cite{Gusynin},
\begin{align}
&\sigma_0(\omega)=\frac{e^2 v_{F}^2|e B|(\hbar \omega +2i\Gamma)}{\pi c i}\times \displaystyle\sum\limits_{n=0}^{\infty} \\ \nonumber
&\left[ \frac{[n_F(M_{n})-n_F(M_{n+1})]-[n_F(-M_{n})-n_F(-M_{n+1})]}{(M_{n+1}-M_n)^3-(\hbar \omega +2i \Gamma)^2(M_{n+1}-M_n)}+\right. \\ \nonumber
&\left. \frac{[n_F(-M_{n})-n_F(M_{n+1})]-[n_F(M_{n})-n_F(-M_{n+1})]}{(M_{n+1}+M_n)^3-(\hbar \omega +2i \Gamma)^2(M_{n+1}+M_n)}\right],
\end{align}
\begin{align}
&\sigma_H(\omega)=-\frac{e^2 v_{F}^2e B}{\pi c}\times \displaystyle\sum\limits_{n=0}^{\infty} \\ \nonumber
&\left([n_F(M_{n})-n_F(M_{n+1})]+[n_F(-M_{n})-n_F(-M_{n+1})] \right)\times \\ \nonumber
&\frac{2(M_n^2+M_{n+1}^2-(\hbar \omega +2i\Gamma)^2)}{(M_n^2+M_{n+1}^2-(\hbar \omega +2i\Gamma)^2)^2-4M_n^2M_{n+1}^2},
\end{align}
where $M_n$ is the energy of the corresponding  Landau level, $M_n=v_F\left({2\hbar e B}/c\right)^{1/2}$, $n_F$ is the Fermi-Dirac distribution function, $v_F$ is the Fermi velocity of the electrons in graphene, and $\Gamma$ is the electron scattering rate. For the numerical studies of the surface wave dispersion, we use the chemical potential $\mu$ equal to $44$ meV and temperature $T=10$ K. Electron scattering rate $\Gamma$ is chosen to be $1.3$ meV which is in agreement with experimental results~\cite{Natexp}. In this case Fermi energy lies between the zero and the first Landau levels. We consider the frequency range from 2 meV to 300 meV. There exist three allowed transitions between the Landau levels in this frequency range which are depicted in Fig.~\ref{fig2}(a).

Transition from zero to the first Landau level corresponds to low-frequency resonance in longitudinal and Hall conductivities [see Figs.~\ref{fig2}(b,c)]. Transitions from the $-2$nd to the $1$st and from the $-1$st to the $2$nd Landau levels have the same frequency and correspond to the high frequency resonance in conductivity dispersions.  Dispersion of the longitudinal and Hall conductivities are shown in Figs.~\ref{fig2}(b,c). Resonances in the dispersion of conductivities associated with the transitions between the corresponding Landau levels can be controlled with the magnetic field.

To demonstrate the importance of the Hall conductivity in the presence of an external magnetic field,
 we compare the obtained dispersion relations with the case where we only account for the longitudinal part of the conductivity. Numerical calculations are performed for the values of $\varepsilon_1=4$ and $\varepsilon_2=1$. Dispersion of the real and imaginary parts of the waveguide number $\beta$ are presented in Figs.~\ref{fig2}(d,e).

When the magnetic field is absent, only one surface wave can exist for each frequency: the TE-like polarized mode, for the case when $\mathrm{Im}(\sigma_0)<0$, and TM-like polarized mode, for the case when $\mathrm{Im}(\sigma_0)>0$. Hall conductivity significantly  changes the real and imaginary parts of waveguide numbers and corresponding properties of surface waves, so that the surface waves become hybrid with TE and TM polarizations mixed.

\begin{figure}[!h]
\centerline{\includegraphics[width=0.9\columnwidth]{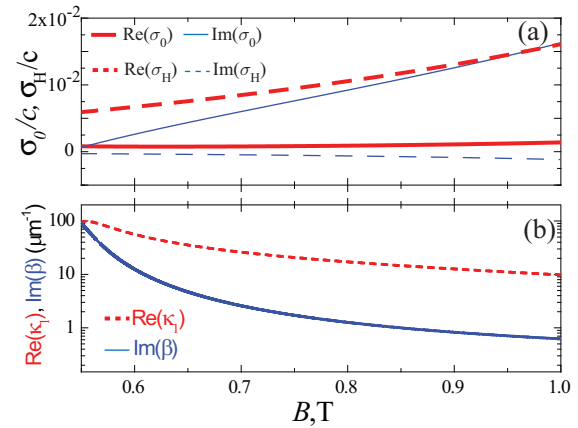}}
\caption{(a) Dependence of the longitudinal and Hall conductivity of graphene on the external magnetic field; the frequency is 0.19 eV. Red lines correspond to a real part of conductivity, blue -  to an imaginary part; solid lines correspond to longitudinal conductivity, dashed - to Hall conductivity. (b) Dependence of the inverse propagation length (solid blue line) and inverse localization length (red dashed line) on the external magnetic field. Frequency is 0.19 eV.  }
\label{fig4}
\end{figure}

Dependence of the surface-wave dispersion on the external magnetic field suggests a new degree of freedom for tuning the surface modes in graphene. To illustrate this property, in Fig.~\ref{fig4} we plot the dependence of the waveguide number on the value of the external magnetic field at the fixed frequency $\omega=0.19 eV$. As shown in Fig.~\ref{fig4}(b), changing the magnetic field from 0.5 to 1 Tesla we can change the localization length of the surface waves, which is inversely proportional to $\mathrm{Re}(\kappa_1)$, and the propagation distance, which is inversely  proportional to $\mathrm{Im}(\beta)$, by 
at least two orders of magnitude.

In conclusion, we have demonstrated that applying external magnetic field to a single layer of graphene place on a dielectric substrate can significantly modify its optical properties. We have shown that in the presence of magnetic field TE-TM hybrid surface plasmons can propagate along the graphene-covered surface, and that the properties of these surface waves can be controlled by changing the external magnetic field.

\end{document}